\documentstyle[floats,epsfig,prd,multicol,aps]{revtex}

\begin{document}

\draft

\title{No time machine construction in open $2+1$ gravity with timelike
 total energy momentum.}
\author {Manuel H. Tiglio\thanks{Electronic address: tiglio@fis.uncor.edu} }
\address{\it Facultad de Matem\'{a}tica, Astronom\'{\i}a y F\'{\i}sica,
Universidad Nacional de C\'{o}rdoba \\ Ciudad Universitaria \\
5000 C\'{o}rdoba, Argentina. }
\maketitle

\begin{abstract}
It is shown that in $2+1$ dimensional gravity an open spacetime with
 timelike sources and total energy momentum cannot have a stable compactly
 generated Cauchy horizon. This constitutes a proof of a version of Kabat's
 conjecture and shows, in particular, that not only a Gott time machine
 cannot be formed from processes such as the decay of a single cosmic string
 as has been shown by Carroll et al., but that, in a precise sense, a time
 machine cannot be constructed at all.
\end{abstract}


\section{Introduction and Overview}
Partly because of the possibility that topological defects such as cosmic
 strings may have been formed in the early universe, and also because of the fact
 that it had already been noted that some solutions in $2+1$ gravity
 corresponding to spinless particles do not have closed timelike curves
 (ctc) if the total energy momentum (${\cal EM}$) is timelike 
 \cite{deser1,wael}, Gott's solution \cite{gott1} has stimulated work
 discussing whether this spacetime is physically reasonable or not. The
 relation between cosmic strings and $2+1$ particles comes from the property
 that the spacetime of an infinitely long and stationary gauge cosmic string
 asymptotically tends to Minkowski spacetime with a deficit angle
 \cite{garfinkle}, and in the cases of interest the core is small enough
 that one can consider just Minkowski with a conical singularity (none of
 these properties holds for gauge but supermassive \cite{garlaguna} or
 global strings \cite{global}). Thus, Gott's solution approximates the
 spacetime of two infinitely long parallel gauge cosmic strings, but it can
 also be thought of as the spacetime of two (spinless) particles in $2+1$.
The first objections to Gott's spacetime were due to the belief that it did
 not have an associated initial value problem, and to the fact that its
 total ${\cal EM}$ is timelike, in some similarity with tachyons
 \cite{deser2}. Approximately at the same time, Cutler showed that in Gott's
 spacetime there are regions without ctc, in particular in these regions
 there are achronal, edgeless, not asymptotically null surfaces, so that it
 can be  thought that the spacetime evolves from an initial data in any of
 these surfaces \cite{cutler} (these surfaces must be suitably chosen, in
 this sense see also \cite{ori}). The apparent analogy with tachyons comes
 from the fact that parallel transport of vectors around a  Gott pair is the
 same as for a tachyon, but this is not true for parallel transport of
 spinors \cite{carroll1}, basically because a Gott pair satisfies the
 dominant energy condition (also the weak and strong ones) while a tachyon
 does not. Therefore, spatial ${\cal EM}$ must not necessarily be considered
 as unphysical (for more discussions on ctc in $2+1$ gravity and on the non
 tachyonic character of a Gott pair, see \cite{gott2}). But then it remains
 intriguing that all known exact solutions describing spinless particles do
 not have ctc if their total ${\cal EM}$ is timelike 
 \cite{deser1,wael,gott1}. Kabat has suggested that this is a general
 feature, specifically,  that spacetimes with spinless particles and
 timelike total ${\cal EM}$ do not have ctc \cite{kabat}. To this we should
 add that 't Hooft has shown that although a Gott pair can be produced from
 initial data with timelike ${\cal EM}$ momentum in a compact surface, a
 `crunch' will occur before the appearance of ctc \cite{hooft}.

Time machine constructions have been associated with compactly generated
 Cauchy horizons (CGCH) \cite{wald3,hawk2}. This is, on one side, because if
 for certain initial data on a surface ${\cal S}$ a domain of dependence
 without a Cauchy horizon is obtained, and changing the data in a compact
 region of ${\cal S}$ a Cauchy horizon appears, beyond which ctc exist, then
 it is compactly generated. On the other side, in certain points of a CGCH
 (the so called base points) strong causality is violated. In this work we
 will follow this approach and take the question of whether a time machine
 can be constructed in an $2+1$ open spacetime with timelike total ${\cal
 EM}$ as equivalent to asking whether such a spacetime can have a CGCH. The
 answer will be negative.

Note that working with a CGCH we get rid of a difficulty present in other
 formulations of Kabat's conjecture. This arises from the fact that it is
 not a priori obvious that, in a spacetime with ctc, a foliation in surfaces
 in which `matter contributes positively' exists, so that one can calculate
 the total ${\cal EM}$ via holonomy, without `counting matter more than
 once'. Specifically: we are interested in spacetimes arising from initial
 data, i.e., of the form  ${\cal D}^+\left( {\cal S} \right)$ where ${\cal
 S}$ is a simply connected, non compact, closed, achronal and edgeless
 surface and its future domain of dependence (a stably causal region) is
 denoted by  ${\cal D}^+\left( {\cal S} \right)$. The dominant energy
 condition, i.e., that $T_{ab} t^a $ is a future directed timelike or null
 vector for all future directed timelike or null $t^a$, choosing $t^a$ as
 the normal to ${\cal S}$, ensures that total ${\cal EM}$ is independent of
 time (a conserved quantity) and of the foliation. If there exists a Cauchy
 horizon, ${\cal H}^+\left( {\cal S} \right)$, then the definition can be
 extended to the horizon if the matter `crosses it', e.g., assuming that
 there are no lightlike sources; specifically, that $T_{ab} t^b  $ is future
 directed and timelike for all future directed timelike $t^a$ (DECa). In
 this work we will assume this energy condition but without requiring that
 $T_{ab} t^b  $ is future directed (DECb), since in a CGCH the weak energy
 condition (WEC) is violated \cite{hawk2}.

There are some previous results in connection with Kabat's conjecture:
 Seminara and Menotti have shown, assuming additional rotational symmetry
 and the WEC, that if there are no ctc at infinity then there are no ctc at
 all \cite{menotti1}; Headrick and Gott have shown that if a ctc is
 deformable to infinity, then the holonomy of the ctc itself cannot be
 timelike, except for a rotation of $2\pi$ \cite{gott2}. Nevertheless, in a
 noncompact CGCH the WEC is violated and the first result is not related to
 time machine construction in the sense made precise above, while the total
 holonomy of a ctc is in principle not related to the total ${\cal EM}$, due
 to the problem mentioned in the preceding paragraph.

In Section II and Section III we will summarize and discuss some known
 results that are crucial in our proof, which will be given in Section IV.

\section{Compactly generated Cauchy horizons}

Since in Section IV we will analyze the dynamics of a CGCH in $2+1$, we need
 here to recall some properties of Cauchy horizons, obtained in $3+1$
 gravity, but equally applicable to $2+1$ gravity. Let ${\cal S}$ be a (partial) Cauchy surface for
 ${\cal D}^+\left( {\cal S} \right) $, an orientable, time orientable
 spacetime with a future Cauchy horizon ${\cal H}^+ \left( {\cal S} \right)
 $. Then:

\begin{enumerate}
\item ${\cal H}^+\left( {\cal S} \right)$ is compact iff ${\cal S}$ is
 compact (see, for instance, \cite{wald1,hawk1}).
\item ${\cal H}^+\left( {\cal S} \right)$ is differentiable everywhere
 except in a set of zero measure. We will assume implicitly
 differentiability of the horizon each time it is needed. That is, we will
 assume, e.g., that the set of non differentiability is not dense (in this
 sense, see \cite{chrusciel1}). 
\item ${\cal H}^+\left( {\cal S} \right)$ is generated by null geodesics
 that are complete in the past but may be incomplete in the future  \cite{wald1,hawk1}. Let us
 denote them, generically, by $\beta(s, x) : {\cal I} \times {\cal
 H}^+\left( {\cal S} \right)  \rightarrow {\cal H}^+\left( {\cal S}
 \right)$, with ${\cal I}$ some interval of ${\cal R}$ and $s$ some affine
 parameter and, unless otherwise stated, we always refer to generators
 directed to the past.
\item ${\cal H}^+ \left( {\cal S} \right)$ is defined as compactly generated
 if all these geodesics enter some compact, connected region ${\cal K}$ and
 remain there forever. That is, for each $x \in {\cal H}^+ \left( {\cal S}
 \right)$, there exists $s_0$ such that $\beta(s,x) \in {\cal K}$ for $s
 \geq s_0$ \cite{wald3,hawk2}.
\item In a non compact CGCH the WEC is violated, i.e., there exist points in
 ${\cal K}$ in which $T_{ab} k^a k^b < 0$ , with $k^a$ the tangent to the generators \cite{hawk2}.
\item The base set, ${\cal B} \in {\cal H} ^+ \left( {\cal S} \right)$, is
 defined as the set of terminal accumulation points of null generators. It
 can be seen that ${\cal B}$ is non empty (this follows from the completeness
 of the generators in the past and the compactness of ${\cal K}$), that
 strong causality is violated in ${\cal B}$ (also from the completeness of
 the generators in the past), and that ${\cal B}$ is comprised by future and
 past inextendible null generators contained in ${\cal B}$, although not
 necessarily closed (`fountains') \cite{wald3} (as we will see in the last
 section, the last statement does not hold in $2+1$).
\end{enumerate}

\section{Total Mass in $2+1$}

   From now on we will assume that the spacetime is open and that the total
 ${\cal EM}$ is timelike. Carroll et al. have shown that in such spacetimes,
 if they are composed of (spinless) particles, there cannot exist any
 subsystem with spatial ${\cal EM}$ \cite{carroll1}. In particular, a Gott
 time machine cannot be created out of the decay of a single cosmic string
 because there is not enough energy for that \cite{carroll2}. In principle
 this property is not obtained as a partial result in the version of Kabat's
 conjecture that we prove here, since a Gott pair satisfies the WEC and,
 indeed, it can explicitly be seen that it does not have a CGCH
 \cite{cutler}. We mention it because a slight generalization is crucial in
 our proof. So, we need here to summarize the analysis given in
 \cite{carroll1}.

Suppose, then, that the matter is composed by particles (assuming implicitly,
 in this way, the DECa). The total ${\cal EM}$ as defined by holonomies is
 constructed starting from a trivial loop in ${\cal S}$ (at constant but
 arbitrary time) and deforming it until it encircles all the particles. In
 the process, the corresponding holonomic operator describes a curve (let us
 call it $\gamma$) in the Lie group , which starts at the indentity
 (corresponding to the trivial loop) and finishes at the total ${\cal EM}$.
We remark that (up to similarity transformations) the total ${\cal EM}$
does not depend on the way in which
 the deformation is carried out, although $\gamma$ does not share this
 property, and is therefore not unique.

\begin{figure}[h]
\centerline{\psfig{figure=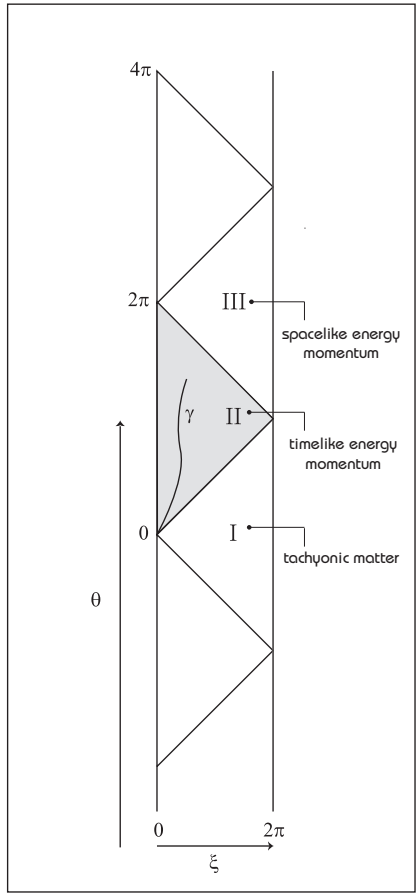,width=10cm}}
\caption{A conformal diagram of the universal covering of ${\cal SO}(2,1)$.}
\label{fig1}
\end{figure}

Coordinates for the double covering of ${\cal SO}(2,1)$, ${\cal SU}(1,1)$,
 can be chosen by decomposing every element in a rotation through angle
 $\theta$ followed by a boost of rapidity $\zeta$ and direction defined by
 the polar angle $(\psi + \theta )/2$. In this coordinates, the metric of
 ${\cal SU}(1,1)$ (naturally given by the structure constants) is
\begin{equation}
ds^2 = -\frac{1}{4} \cosh{ \frac{\zeta }{2} } d\theta ^2 + \frac{1}{4}
 d\zeta ^2 + \frac{1}{4} \sinh{ \frac{\zeta }{2} } d\psi ^2 \label{metrica}
\end{equation}
which shows that ${\cal SU}(1,1)$ has the geometry of anti-de Sitter
 spacetime. A conformal diagram of (the universal covering of) this
 spacetime is shown in Fig.\ \ref{fig1}, with one dimension suppressed and
 $\xi \equiv 4 \tan ^{-1}{(e^{\zeta /2})} - \pi$. Systems with timelike
 (spacelike) total ${\cal EM}$ lie in region II (III).

Since we have assumed that the total ${\cal EM}$ is timelike, the
 corresponding holonomic operator is equivalent (through a similarity
 transformation) to a rotation through a certain angle $\theta _{total}$,
 which is defined as the total mass. Since the topology of ${\cal S}$ is
 assumed to be ${\cal R}^2$ and (it is assumed that) the initial data are
 geodesically complete,
\begin{equation}
\theta _{total} = \int_{\Sigma} K dA  \leq 2 \pi       \label{bonete}
\end{equation}
where $K$ is the gaussian curvature, the equality and inequality follow from
 the Gauss - Bonnet and Cohn - Vossen theorems, respectively.

At this point we remark that the matter need not be composed of particles
 (if it is not, the loop must encircle the support of $T_{ab}$ or be
 deformed to infinity if this support is not compact) and that the same
 analysis holds if one just assumes DECa. This condition implies that
 $\gamma$ is timelike and future directed, a condition that in turn implies
 that no subsystem can have spacelike ${\cal EM}$; since, if $\gamma$
 crosses from region II to region III, it cannot return back to region II and
 lead to timelike total ${\cal EM}$, as we did assume (see \cite{carroll1} for further details).

We should also remember that Seminara and Menotti  have explicitly shown
 (without using the causal structure of anti-de Sitter spacetime, or the
 Gauss - Bonnet and Cohn - Vossen theorems) similar properties
 \cite{menotti2}. Specifically, that in an open spacetime with timelike
 total ${\cal EM}$ satisfying DECa the mass increases as the loop encircles
 more and more matter and if, having reached the total mass of $2\pi$, more
 matter is encircled then the total ${\cal EM}$ turns null or spacelike.

Now recall that there is already a `standard' formulation for asymptotic
 flatness in $2+1$, with analogues to the ADM \cite{ash1} and Bondi masses
 \cite{ash2}. In the hamiltonian formulation it is required that 
 asymptotically the spacetime approaches that of a (spinless) particle,
 i.e., Minkowski with a deficit angle: this angle defines the `ADM' mass.
 Note that for such a spacetime the total holonomy of a loop that is
 deformed to infinity is equivalent to a rotation through an angle which
 coincides with the deficit angle, so the `ADM' mass coincides with that
 defined via holonomies. It was emphasized in \cite{ash1} that in $2+1$, in
 contrast to $3+1$, the total mass is not only bounded from below but also
 from above. This was noted in a hamiltonian formulation, in which case the
 total mass must be strictly less than $2\pi$ and it was argued that it is
 not a feature arising from such formulation analyzing the limit of a
 particle's spacetime when the mass approaches and subsequently exceeds
 $2\pi$, in which case the conical structure becomes cylindrical and
 subsequently geodesically incomplete. What we want to remark is that, on
 account of Eq. (\ref{bonete}), and the discussion that preceded it, if
 geodesically completeness is assumed, then the total mass is effectively
 bounded, it must be $\leq 2\pi$. Let us also remark that not every compact
 system is asymptotically flat in the sense described in \cite{ash2}, since
 a spacetime with spacelike or null total ${\cal EM}$ is not, even if it has
 $T_{ab}$ with compact support and thus curvature of compact support (the
 Weyl tensor vanishes identically in $2+1$).

\begin{figure}[h]
\centerline{\psfig{figure=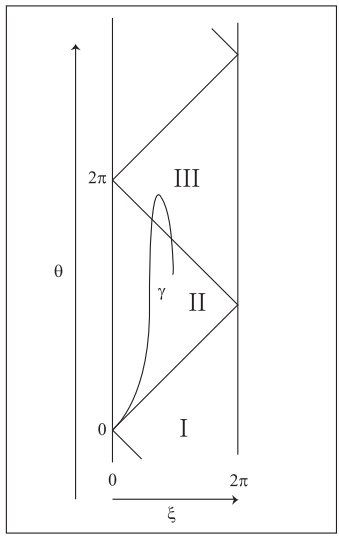,width=6cm}}
\caption{When the WEC is violated $\gamma$ is past directed, and can cross
 from region II to region III and turn back to region II, having thus timelike
 total ${\cal EM}$ and subsystems with spacelike ${\cal EM}$.}
\label{fig2}
\end{figure}

Let us assume now that DECb holds. Then $\gamma$ is timelike but not
 necessarily future directed and there are some subtleties that we need to
 discuss. Firstly, we shall assume that the total mass is non negative.
 Therefore, near the identity $\gamma$ can be chosen non past directed by
 simply choosing the point where the loop is initially expanded as one in
 which there is non negative mass. Now note that without assuming DECa the
 reasoning that led to the non existence of subsystems with spacelike
 ${\cal EM}$ is, in principle, no longer true (see Fig.\ \ref{fig2}). We
 shall overcome this difficulty by imposing that, once that $\gamma$ has
 been chosen non past directed at the identity, it remains in region II.

\section{The Proof}

Considering the previous sections, the proof of the version of Kabat's
 conjecture we give here reduces to showing that open $2+1$ spacetimes with
 a CGCH must have a subsystem with null or spacelike ${\cal EM}$, or
 timelike ${\cal EM}$ of mass zero or $2\pi$. Imposing timelike total ${\cal
 EM}$ and the DECb, we will have arrived at a contradiction, except for the
 last two cases, which will turn out to be unstable, in an appropriate sense.
 Such a subsystem is, precisely, that encircled by the closed null geodesic
 whose existence we will now prove.

So, in $2+1$, not only strong but also stable causality is violated: the
 base set necessarily contains at least one closed null geodesic ${\cal C}$.
 The proof follows from the Poincar\'e - Bendixon - Schwartz
 theorem \cite{sch} applied to the dynamical system defined by the past
 directed null generators $\beta (s,x)$ in the two - dimensional compact
 manifold ${\cal K}$. We start by noting some properties of this dynamical
 system.:

\begin{enumerate}
\item ${\cal K}$ is positively invariant (this results from the
 definition of ${\cal K}$).
\item The past directed generators $\beta (s,x)$ in ${\cal K}$ exist globally
 (they are complete in the past, and we are always referring to this
 direction in time).
\item From completeness of the generators in the past, there are no fixed
 points, i.e., there does not exist $x \in {\cal K}$ such that $\beta(s,x) 
= x \;\; \forall s$.
\end{enumerate}

The first item allows us to think of our dynamical system as one in a
 compact manifold. The second item allows us to introduce what is usually
 called the `$\omega$ limit set' of a point $m \in {\cal K}$. This set is defined as the set of points $x \in {\cal K}$ such that the generator
 passing through $m$ satisfies: for every open neighborhood ${\cal O}$ of
 $x$ and every $s_0$ (in the domain in which they are defined) there exists
 $s > s_0$ such that $\beta (s,m) \in {\cal O}$. The `$\omega$ limit set of
 ${\cal K}$' is, similarly, defined as the union of the $\omega$ limit sets
 of all $y \in {\cal K}$. That is, the $\omega$ limit set of ${\cal K}$ is
 {\em by definition} the base set ${\cal B}$. With this in mind, we shall
 replace $\omega$ by ${\cal B}$.

The Poincar\'e - Bendixon - Schwartz theorem shows the following: let ${\cal
 K}$ be a compact, connected, orientable two dimensional manifold with $k^a
 \in T({\cal K})$ and complete orbits, such that, for $m \in {\cal K}$, ${\cal
 B}(m)$ contains no fixed points (our dynamical system does satisfy these
 conditions). Then either
\begin{itemize}
\item ${\cal B}(m) = {\cal K} = {\cal T}^2$; or
\item ${\cal B}(m)$ is a closed orbit ${\cal C}$, and $\beta (s,m)$ winds
 towards ${\cal C}$,
\end{itemize}
where ${\cal T}^2$ is the torus, a manifold without boundary, excluded in our case. Thus we have a closed null geodesic ${\cal C}$.

In the proof of the Poincar\'{e} - Bendixon - Schwartz theorem the
 2-dimensionality of ${\cal K}$ is crucial. Indeed, although not related to
 the failure of this theorem or the properties of CGCH in $2+1$ gravity, it
 has been emphasized that (in $3+1$) the base points are not, in general,
 made up by closed null geodesics (`fountains') \cite{chrusciel2}.

 We will now show that the ${\cal EM}$ encircled by ${\cal C}$ is spatial,
 null or timelike of mass zero or $2\pi$. For that purpose consider an
 arbitrary base point $p \in {\cal C}$, and an orthonormal base $\{ v^a \}
 \in T_p$. Parallel transport of such a base around the loop ${\cal C}$
 defines a new orthonormal base $\{ V^a \} \in T_p$, related to the previous
 one by a proper Lorentz transformation $L:\{ v^a \} \rightarrow \{ V^a \}$,
 whose equivalence class defines the ${\cal EM}$ encircled by ${\cal C}$. 
 In particular, let us take a base such that the tangent vector to ${\cal
 C}$ at $p$, $k^a$, belongs to $\{ v^a \}$. ${\cal C}$ being a geodesic,
 $k^a$ is parallel transported and, since it is a non broken geodesic, $K^a$
 (defined by $L k^a = K^a$) is proportional to $k^a$, i.e., $K^a =
 \lambda k^a$. In other words, $k^a$ is an eigenvector of $L$. It is
 straightforward to see that if $L$ is timelike it has no null eigenvector,
 except when $L$ is the identity, in which case the eigenvalue is obviously
 $1$ and all the vectors are eigenvectors. When $L$ is null it has exactly
 one null eigenvector, with eigenvalue $1$ and when it is spatial it has two
 null eigenvectors, with eigenvalues $\lambda_1 >1$ and $\lambda_2 = 1/
 \lambda_1$, so the first is attractive and the second repulsive. Then, the
 situation is different from $3+1$, because in that case $L$ defines a map
 on the (past) sphere of null directions, and every orientation preserving
 map on the sphere has at least one fixed point, so that in $3+1$ every
 proper Lorentz transformation has at least one fixed null direction
 \cite{penrose}. On the other hand, a homeomorphism on the circle which
 preserves orientation has a fixed point iff its {\em rotation number} is
 zero \cite{devaney}; in particular for rotations this number coincides with
 the angle of rotation, so one recovers that if $L$ is timelike it has null eigenvectors iff it is the identity.

The point is that, applying the analysis of the previous paragraph to the
 closed null geodesic ${\cal C}$, we have shown that the ${\cal EM}$ it
 encircles is spatial, null or timelike of mass zero or $2 \pi$. Although it
 is not related to our proof, remember that the eigenvalue cannot be $ < 1$
 because otherwise it can be shown that there would be ctc in ${\cal
 D}^+\left(  {\cal S} \right)$  \cite{hawk2} (see also Proposition 6.4.4 of
 \cite{hawk1}), a stably causal region. So, when the ${\cal EM}$ encircled
 by ${\cal C}$ is spacelike the corresponding eigenvalue is $\lambda  =
 \lambda _2> 1$. 

\begin{figure}[h]
\centerline{\psfig{figure=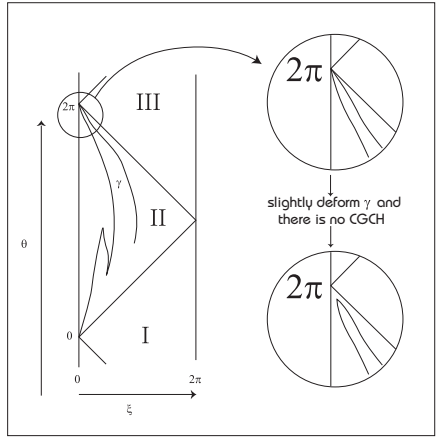,width=6cm}}
\caption{This shows an example where a subsystem has $2\pi$ mass, but
 $\gamma$ can be slightly deformed to rule out this possibility.}
\label{fig3}
\end{figure}

Summarizing, if the total ${\cal EM}$ is timelike and DECb holds, there are
 no subsystems with null or spacelike ${\cal EM}$. Thereby, the ${\cal EM}$
 encircled by ${\cal C}$ can only be timelike of mass zero or $2 \pi$. If we
 had supposed DECa we would have been able to discard the case of zero mass
 because ${\cal C}$ would encircle a simply connected flat (vacuum $+$ the
 identical vanishing of the Weyl tensor) region and thus causally well
 behaved. However, since WEC is violated, zero mass does not necessarily
 correspond to vacuum. Nevertheless, both cases, subsystems with zero or
 $2\pi$ mass, are unstable, in the sense that in every neighborhood (with
 the Lie group manifold topology) of these points all timelike elements do
 not have fixed null directions (these two points belong to the boundary of
 region II). By slightly altering the distribution of masses (or $\gamma$,
 equivalently) there will be no subsystem with $0$ or $2\pi$ mass, see Fig.\
 \ref{fig3}.

\section{Final Remarks}

Carroll et. al have shown, from energy considerations, that a Gott time
 machine cannot be constructed in $2+1$ dimensional open gravity with
 timelike sources and total energy momentum. In this paper we have shown
 that, in a precise sense, a time machine cannot be constructed at all,
 providing a proof of a suitable version of Kabat's conjecture.

Note that it makes sense to talk about total ${\cal EM}$ of a time machine:
 this quantity remains unalterated if the initial data is changed in a
 compact region of ${\cal S}$, since it is defined by a loop that encircles
 (in particular) such a region.

The proof is constructive in some aspects, e.g., it shows the existence of
 closed null geodesics in a CGCH, a property which is interesting in its
 own.

In a non compact CGCH the WEC is violated, the known classical fields obey
 this condition but they do not when quantized (even in Minkowski), although
 an averaged version, the averaged null energy condition, has been proven to
 hold in some cases \cite{wald4}. Therefore, it can be said that in order to
 create a time machine quantum matter is needed, and it is natural to ask
 whether the laws of physics allow ctc or a CGCH. There have been different
 and opposite conclusions to this question (see, for instance,
 \cite{hawk2,thorne}) and it seems reasonable to say that it will be
 difficult to have a complete answer within semiclassical gravity since not
 even the usual quantum field theory can be extended from ${\cal D}^+\left(
 {\cal S} \right)$ to the base set of a CGCH \cite{wald3}.

\section*{Acknowledgements}
The author would like to thank R. J. Gleiser, J. Pullin, and G. A. Raggio
 for encouraging this work and helpful suggestions which improved its
 presentation, and CONICOR for financial support. This work was supported in
 part by grants from the National University of C\'ordoba, and from CONICOR,
 and CONICET (Argentina).

\end{document}